\documentstyle[aps,epsfig]{revtex}  % Specifies the document style.
\newcommand{\be}{\begin{equation}}
\newcommand{\ee}{\end{equation}}

\begin{document}
\title{Renormalization Group for Large $N$ Strongly Commensurate Dirty Boson Model}
\author{M. B. Hastings}
\address{Physics Department, Jadwin Hall\\
Princeton, NJ 08544\\
hastings@feynman.princeton.edu}
\maketitle
\begin{abstract}
The large $N$ sigma model, in $D<4$ space-time
dimensions, with disorder a function
of $d$ space dimensions, is analyzed via a renormalization group treatment.
Critical exponents for average quantities are calculated, first to lowest order
and then to all orders, in 
$\epsilon=D-2-\frac{d}{2}$.
In particular,
it is found that $\nu d =2$.  When $D=d+1$, this
model is equivalent to a large $N$ limit of the strongly commensurate
dirty boson problem.
   \end{abstract}
\section{Introduction}
The problem of the statistical mechanics of a system with quenched
randomness is very interesting and difficult, combining the problems of
disorder and interaction.  We can consider two cases, the ordinary
statistical mechanics of such a system, or the quantum statistical mechanics,
in which a $d$ dimensional quantum system can be considered as a $d+1$
dimensional classical system, with disorder correlated in the time direction.
These correlations make the effects of disorder much stronger, as seen,
for example, in the one dimensional disordered quantum Ising chain\cite{fisher},
a problem which is probably the most well understood disordered quantum phase
transition.  There has also been much work on the quantum
statistical mechanics of a system of repulsively interacting 
bosons in a disordered potential,
known as the dirty boson problem\cite{dirtyboson}, but this problem
is not as well understood.

An action that may be used to describe the dirty boson problem is
\be
\int d^dx \, dt \, \Bigl( 
-\partial_{x} \overline \phi(x,t) \partial^{x} \phi(x,t)
-\partial_{t} \overline \phi(x,t) \partial^{t} \phi(x,t)
+w(x)\overline\phi(x,t)\partial_t\phi(x,t) -
U(x) \overline\phi\phi +\frac{g}{N}(\overline\phi\phi)^2 \Bigr)
\ee
Here, $U(x)$ and $w(x)$ are considered to be quenched random variables, with small
fluctuations.
Although the action above describes a $D=d+1$ dimensional system,
the randomness in $U(x)$ and $w(x)$ is a function 
of the $d$ dimensions $x$ only.
This action is expected to describe a system with a number of different
phases.  If the average chemical potential $U$ is negative, the system is
a Mott insulator, which has a gap.  
Increasing the chemical potential, the system then becomes
a Bose glass, and finally a superfluid.  The Bose glass is characterized
by a gapless phase with infinite superfluid susceptibility\cite{boseglass}.
In this phase, the bosons occupy localized states, with a localization length
that diverges as the superfluid transition is approached.

If the boson density commensurates with a lattice, then $w(x)$ vanishes on
average; this can be accomplished experimentally by tuning a chemical
potential.  In a typical experimental situation, however, there will
still be fluctuations in $w(x)$.  This case, with nonvanishing fluctuations
in $w(x)$, is the weakly commensurate case.  If $w(x)=0$ identically, then
we have the strongly commensurate case.  In an experimental setup with
either a dirty boson system or a Josephson junction array, this would require
tuning the local chemical potential to be constant, so there is no local
breaking of particle-hole symmetry due to disorder,
while still having some fluctuations in the hopping term.  This is
obviously a more difficult task.

In this paper, we consider only the strongly commensurate case, where $w(x)=0$
everywhere.  Although less physically applicable to boson systems,
we feel that the results on the strongly commensurate case are interesting
in themselves, as well as being applicable to some other problems of quantum
critical points.  In particular,
there exist quantum critical systems, with an $O(3)$ order parameter, which
in the disordered case may be well described by an $N=3$ disordered rotor model
like those consider in this paper.  See work by Sachdev\cite{sachdev}, and
references therein to collinear quantum antiferromagnets.

In the strongly commensurate case, we will not see the Bose glass phase.
As the chemical potential is increased, the system will go from a gapped 
Mott insulator phase to a gapless Griffiths phase, but the 
superfluid susceptibility will remain
finite.  One reason for considering the strongly commensurate case is
related to the use of the large $N$ limit and will be discussed later.  Results
will be described in a future work for the weakly commensurate case\cite{me2}.
However, we feel that it is useful, in this paper, to mention the existence 
of the weakly commensurate problem, and the Bose glass found in that problem, 
to understand the different phase diagram with and without the $w(x)$ term.
The Bose glass phase, as we understand it, relies on the existence of the
$w(x)$ term, and so, contrary to what other authors suggest\cite{xiaogang},
we feel it is impossible to have such a phase in the strongly commensurate 
problem.  Since the problem without the $w(x)$ term can be understood simply
as a rotor problem, it seems that the only possible phases are paramagnetic
and ferromagnetic, with a Griffiths phase near the critical point.

In this paper, we consider a large $N$ generalization of the strongly
commensurate dirty boson problem\cite{xiaogang}.  We consider a system
described by the following partition function
\be
\int \Bigl(\prod\limits_{x,t}\delta(\overline\phi(x,t)\phi_i(x,t)-N\sigma^2(x))\Bigr)
[d\phi_i] e^{-\int d^dx \, d^{(D-d)}t
(\partial_{x} \overline \phi_i(x,t) \partial^{x} \phi^i(x,t)+
\partial_{t} \overline \phi_i(x,t) \partial^{t} \phi^i(x,t))
}
\ee
where the field $\phi_i$ has $N$ components, and is a function of
$d$ space coordinates $x$ and $D-d$ time coordinates $t$.
The $\delta$-function constraint on the length of the spins is
technically simpler and is not expected to 
alter the universality class from the quartic interaction
considered above.  The function $\sigma^2(x)$ is a function of the spatial
coordinate $x$ only, and is related to the spatially varying $U(x)$ considered
in the previous equation.

The advantage of the large $N$ formulation of the problem is that the
system may be exactly solved for any fixed realization of disorder by
solving a self-consistency equation.  We may replace the $\delta$-function
constraint by an integral over a Lagrange multiplier field $\lambda(x)$ as
\be
\int
[d\phi_i] [d\lambda] e^{-\int d^dx \,d^{(D-d)}t\, \partial_{\mu} \overline \phi(x,t) \partial^{\mu} \phi(x,t) +\int d^dx \,d^{(D-d)}t\,\lambda(x) (\overline \phi \phi -\sigma^2(x))}
\ee
where the integral for $\lambda(x)$ extends from $-i\infty$ to $+i\infty$.
In the large $N$ limit, we can use a saddle point approximation for
$\lambda$, the saddle point being found by the self-consistency equation
\be
\label{sc}
\sigma^2(x)=\langle x,t=0|(-\partial_{\mu}^2 +\lambda(x))^{-1}|x,t=0\rangle
\ee
This equals
\be
\sigma^2(x)=\int d^{(D-d)}\omega \, \langle x,t=0|\Bigl(-\partial_{x}^2 
+\omega^2 +\lambda(x)\Bigr)^{-1}|x,t=0\rangle
\ee
We will write $\sigma^2=\sigma^2_0+\delta\sigma^2$, where $\sigma^2_0$
is a constant, which is tuned to drive the system through the phase
transition, and $\delta\sigma^2$ is a random term.  For small
$\sigma^2_0$ the system is in the Mott insulator phase.
For large $\sigma^2_0$ the system is in the superfluid phase.

In the above equations, we assume that the Green's function on the right-hand
side has been renormalized by subtracting a divergent quantity.  That is, we
will take a Pauli-Villars regularization for the Green's function, and take the
regulator mass to be very large, while adding an appropriate divergent constant to
$\sigma^2$ on the left-hand side.  The cutoff for the regulator is completely different
from the cutoff for fluctuations in $\delta\sigma^2$
that will be introduced for the RG of the next section; the cutoff for the
regulator will be much larger than the cutoff for fluctuations in 
$\delta\sigma^2$ and will be unimportant in the RG.

Note that in the large $N$ limit, it makes sense to define the theory
for non-integer $D-d$, by doing the integral over $\omega$.
For finite $N$, such a definition may have trouble,
and indeed in the double dimensional expansion in $4-d$ and $D-d$, there
is some question about expanding around $D-d=0$\cite{doubled}.

One disadvantage of the large $N$ expansion is that it is slightly
difficult to include the terms linear in the time derivative,
necessary to understand the weakly commensurate and incommensurate dirty
boson problem.  Consider the simple problem
\be
\int
[d\phi_i] e^{-\int dt \, \overline \phi_i(t) \partial_t \phi^i(t) -U(x) \overline\phi_i(t)\phi^i(t) +\frac{g}{N}(\overline\phi_i(t+\delta)\phi^i(t))^2}
\ee
where $\delta$ is a small number introduced for point-splitting.
While for $U<0$ everything is correct, for $U>0$ the solution
of the problem via the self-consistency equation becomes ill-defined for
zero temperature, although not for arbitrarily small, non-zero, temperatures.  
The trouble is in the
zero temperature limit of the problem in which we must consider frequencies
$\omega$ arbitrarily close to $0$.  This will be further discussed in future
work\cite{me2}.

Having dropped terms
linear in the time derivative, we do not expect to see a Bose glass phase.
The original argument for the Bose glass phase was based on considering a
system consisting of localized states for the bosons, with hopping between
the localized states being neglected.  It was then shown that for a random
distribution of chemical potentials for each localized state, the spectrum
of excitations would be gapless with constant density of excitations of
low energy.  These excitations would correspond to changing by one the
number of particles in a given localized state.  However, if one considers
a system of localized states with action containing terms quadratic in the
time derivative instead of linear, in the large $N$ limit, the density of
low energy excitations goes to zero.  In the Griffiths phase of the model
we are considering there are states at arbitrarily low energies, but the
density of states vanishes as $e^{-cE^{-d}}$, where $c$ is some constant.

In the RG treatment, we will be considering the phase transition
between the Griffiths phase and the superfluid phase, with
weak randomness in $\sigma^2(x)$.
By the Harris criterion\cite{harris}, weak randomness is irrelevant at the 
pure fixed point for $\nu d >2$ and marginal for $\nu d =2$.  There is also a 
bound that for a stable disordered fixed
point $\nu d \geq 2$\cite{bound}.  For $D<4$, we find that $\nu=1/(D-2)$.  
Thus, there is a range of values of $D$ and $d$ satisfying $d/(D-2)=2$, at 
which the randomness is marginal.  We are able to construct a renormalization 
group near any of these values.  Note that for the quantum statistical 
mechanics of a $d$ dimensional system, where $D=d+1$, we find that the 
disorder is marginal for $d=2$ and relevant for $d>2$.  The renormalization
group is not constructed by expanding down from an upper critical dimension.
Instead, it is constructed by expanding upwards near a range of dimensions such
that $d/(D-2)=2$.  The RG is only valid for $D<4$, as discussed
in the next section.

Within the RG, we will proceed perturbatively, but will obtain exact
results for all the various exponents for average quantities.  One of the
most striking results is that $\nu d=2$.  This implies that the system
saturates the bound discussed above\cite{bound}.  Similar results have
been found for various other phase transitions.  For example, exact
results for the transverse field Ising model in d=1 lead to saturation
of this bound\cite{fisher}, with $\nu=2$.  
Also, numerical simulations on the quantum Ising spin
glass in 3+1 dimensions show this bound\cite{david}.  This may be
a common feature of quantum phase transitions.

The next two sections will develop the renormalization group.  The procedure
will be to take a problem with fluctuations in $\sigma^2$ up to some
cutoff $\Lambda$, and define another problem which includes fluctuations
in $\sigma^2$ only up to a wavevector $\Lambda-\delta\Lambda$,
such that we preserve the low-momentum correlation functions.  In section II
we will perform a one-loop Wilson-Fisher RG.  To extend this
to higher orders would require keeping track of many operators, while
an alternative procedure discussed section III requires
only keeping track of the renormalization of two quantities: $\sigma_0^2$
and $S$, where $S$ is a measure of the strength of fluctuations in
the disorder $\delta\sigma^2$.  
In order to preserve the low-momentum correlation functions,
we will require that the propagator used includes self-energy corrections
due to the high-wavevector fluctuations, and also require a renormalization of
the disorder strength to produce the same low-wavevector fluctuations
in $\lambda$, up to a renormalization of the vertex.  
Requiring the same low-wavevector fluctuations in $\lambda$
makes the procedure consistent, and permits one to continue iterating the RG.
\section{Renormalization Group}
We proceed with a renormalization group acting directly on the 
self-consistency equation.  The goal will be to start with a given problem
which includes high wavevector fluctuations in $\sigma^2$, 
and find a related problem which includes fluctuations in $\sigma^2$ only
up to a lower wavevector.  We will require that the
correlation functions in the related problem are equal, up to rescaling, to
the correlation functions in the original problem averaged over
the high wavevector fluctuations.  This requirement is what will define
the RG.  

At lowest order, considered in this section, this will require that
$\lambda$ is unchanged, up to rescaling, at low wavevectors.
To this order, we can accomplish this goal by defining a new problem in
the same form as the old problem, via a self-consistency equation with
one parameter $\lambda$; to higher orders this Wilson-Fisher procedure will
be more complicated and we will instead use an alternative technique in 
the next section.  To higher orders within the Wilson-Fisher approach one
encounters operators such a spatially fluctuating gradient terms and
$\omega^2$ terms, as well as non-Gaussian and momentum dependent
distributions of the disorder.

In equation (\ref{sc}), we have broken $\sigma^2(x)$ into two pieces:
$\sigma_0^2+\delta\sigma^2(x)$, where $\delta\sigma^2(x)$ has
vanishing mean.  In the perturbative $\epsilon$ expansion
being developed here, we can perturb in $\delta\sigma^2(x)$ as it will have
fluctuations at the critical point which are of order $\epsilon$.

Consider the self-consistency equation (\ref{sc}), and define a
cutoff $\Lambda$ such
$\sigma^2$ only has fluctuations for wavevectors less than $\Lambda$.
Formally, we can
invert equation (\ref{sc}) to obtain $\lambda$ as a function of $\sigma^2$ as
follows.
For small $\delta\sigma^2$, we can expand the right-hand side of (\ref{sc})
as a power series in $\lambda$.  At criticality, where $\lambda$ vanishes
for vanishing $\delta\sigma^2$, we find to lowest order
\be
\delta\sigma^2(p)=\int d^dk\,d^{D-d}\omega\,
\frac{1}{(p-k)^2+\omega^2} \frac{1}{k^2+\omega^2}\lambda(p)
\ee
This can be derived from a diagram similar to a polarization bubble as
shown in figure 1.

This implies that as a formal inverse, to lowest order,
\be
\lambda(p)=c_1 \delta\sigma^2(p) p^{4-D}
\ee
where $c_1= \frac{1}{\pi}^{D/2} \frac{\Gamma(D-2)}
{\Gamma(2-D/2)\Gamma^2(D/2-1)}$.
For small wavevector components
of $\delta\sigma^2$, this formal inverse is ill-behaved, but for high 
wavevector components when $D<4$,
the procedure given will be correct to lowest
order in $\epsilon$.  This is why we need $D<4$ for the RG, as otherwise
the polarization bubble is divergent for large wavevector instead of
small wavevector.

Define a measure $S$ of the strength of disorder, by assuming that 
\be
\label{Sdef}
\langle\delta\sigma^2(p)\delta\sigma^2(q)=(2\pi)^d\delta(p-q)S
\ee
where to
lowest order we may take $\delta\sigma^2$ to be Gaussian distributed.  Then,
we find that at the cutoff $\lambda$ has mean-square $c_1^2\Lambda^{8-2D}
(2\pi)^dS$.  We will define $L=c_1^2\Lambda^{8-2D}S$ to measure fluctuations
in $\lambda$.
The effect of these components in $\lambda$ will be to renormalize
the propagator for the field $\phi$, as well as to renormalize the
vertex used to calculate the scattering of $\phi$ off $\lambda$ and to
calculate $\sigma^2$.  The self-energy is given
by
\be
\Sigma(p,\omega)=\delta\Lambda\int_{k^2=\Lambda^2} d^{d-1}k
\frac{1}{(p+k)^2+\omega^2}L
\ee
See figure 2 for the appropriate diagram.
This is equal to a constant, which may be absorbed into a renormalization of
$\lambda$, plus
\be
-\frac{\delta\Lambda}{\Lambda}(c_2 p^2+c_3 \omega^2)
L
+ ...
\ee
where $c_2= (1-4/d) c_3$ and
$c_3=2\frac{\pi^{d/2}}{\Gamma(d/2)}\Lambda^{d-4}$.
This implies that the propagator is renormalized to
\be
\frac{1}{(1+\frac{\delta\Lambda}{\Lambda}c_2L)p^2+
(1+\frac{\delta\Lambda}{\Lambda}c_3L)
\omega^2}
\ee
while the vertex renormalization, shown in figure 3, is given by
\be
\frac{\delta\Lambda}{\Lambda}c_3
L
\ee
This implies that we can write a new self-consistency equation
\be
\label{nsc}
(1-\frac{\delta\Lambda}{\Lambda}c_3L)
\sigma^2(x)=\int d^{(D-d)}\omega \,\langle x,t=0|
\Bigl(-(1+\frac{\delta\Lambda}{\Lambda}c_2L)
\partial_{x}^2+ 
(1+\frac{\delta\Lambda}{\Lambda}c_3L)
(\omega^2 
+\lambda(x))\Bigr)^{-1})|x,t=0\rangle
\ee
where $\sigma^2$ now has only components at wavevectors less than $\Lambda
-\delta\Lambda$.  Note that by removing the high wavevector components of
$\lambda$ in the vertex renormalization, we reduce the value of $\sigma^2$
on the left-hand side of equation (\ref{nsc}).
We rescale equation (\ref{nsc}) by defining a new scaled $\lambda$, scaling
the integral over $\omega$, and scaling the $d$ space dimensions, to obtain
\be
\label{sscale}
(1+\frac{\delta\Lambda}{\Lambda}(D-2))
(1-\frac{\delta\Lambda}{\Lambda}(c_3-c_2)L)^{1-\frac{D-d}{2}}
\sigma^2(x)=\int d^{(D-d)}\omega \,\langle x,t=0|
\Bigl(-\partial_{x}^2 
+\omega^2 +\lambda(x)\Bigr)^{-1})|x,t=0\rangle
\ee

Assuming that $\sigma=\sigma_0^2+\delta\sigma^2$, with $\delta\sigma^2$ having
a distribution given by equation (\ref{Sdef}), we can obtain
RG equations for $\sigma_0^2$ and $S$ by considering the scaling of
$\sigma$ under equation (\ref{sscale}).  We obtain
\be
\label{sorg}
\frac{d{\rm ln}\sigma_0^2}{d{\rm ln}\Lambda}=D-2-(c_3-c_2)c_1^2\Lambda^{8-2D}S(1-\frac{D-d}{2})
\ee
\be
\frac{1}{2}\frac{d{\rm ln}S}{d{\rm ln}\Lambda}=D-2-(c_3-c_2)c_1^2\Lambda^{8-2D}S
(1-\frac{D-d}{2})
-d/2
\ee

Therefore, there is a fixed point of the RG flow, for 
$D-2-d/2=(c_3-c_2)c_1^2\Lambda^{8-2D}S(1-\frac{D-d}{2})$.  
The fixed point is stable in the $S$ direction.  It is unstable in the
$\sigma_0^2$ direction.  We find $\frac{d{\rm ln}\sigma_0^2}{d{\rm ln}\Lambda}=d/2$ 
at the fixed point, leading to the exponent $\nu=2/d$, which implies that
the bound $\nu d \geq 2$ is saturated.

In fact, equation (\ref{sorg}) is not quite correct.  Due to scaling of the
regulator mass, there is an additional constant term in the change of
$\sigma_0^2$ with respect to ${\rm ln}\Lambda$, or equivalently an extra term
$\frac{C}{\sigma_0^2}$ in the change of ${\rm ln}\sigma_0^2$ with
respect to ${\rm ln}\Lambda$.  Here, $C$
is some non-universal constant.  This additional term only leads to a
non-universal change in the critical value of $\sigma^2_0$, 
and no change in the critical exponents.

We can obtain the dynamic critical exponent by considering the different
scaling of $p^2$ and $\omega^2$ in the propagator.  This difference is
$(c_3-c_2)c_1^2\Lambda^{8-2D}S=\frac{D-2-\frac{d}{2}}{1-\frac{D-d}{2}}$.  
Therefore, 
\be
\label{dce}
z=1+\frac{D-2-\frac{d}{2}}{2-D+d}=
\frac{d/2}{2+d-D}
\ee
The average Green's function, $\langle G(p,\omega)\rangle$,
is 
\be
\frac{1}{p^{\frac{d}{2+d-D}} +\omega^2}
\ee

This result for the dynamic critical exponent differs greatly from that
expected in the weakly commensurate case, where scaling theory predicts
$z=d$\cite{dirtyboson}.  This is not due
to $1/N$ corrections, but rather a big difference between weakly and
strongly commensurate phase transitions\cite{me2}.
\section{Results to All Orders}
Given the simplicity of the result $\nu d=2$, one might suspect that
the exponents for average quantities remain unchanged to all orders.  A very
simple argument shows that this is true.  To obtain results to higher order, it
is useful to use a formulation of the RG other than the Wilson-Fisher RG.
We will proceed as follows.  Consider a theory with quadratic fluctuations in
$\sigma^2$ of some magnitude $S$, with fluctuations ranging up to some wavevector
$\Lambda$, and with given $\sigma_0^2$.  
In this theory calculate the average Green's function at low momentum,
as well as the low momentum fluctuations in $\lambda$,
to some order in perturbation theory.  
The reason for considering fluctuations in $\lambda$ is that this is
what is needed to continue the computation of
the Green's function to higher orders. 
Then, for a theory with cutoff $\Lambda-\delta\Lambda$, determine the 
appropriate value of $S$ and appropriate coefficients of $\partial_x^2$ 
and $\partial_t^2$ in the action, so as to reproduce the given low energy 
behavior to the same order in perturbation theory, with the same
$\lambda$ up to a renormalization of the scattering vertex.

This procedure is a well-established alternative to the Wilson-Fisher
and Callan-Szymanzik techniques.  It is discussed for example in the
classic reference Domb and Green volume 6, as the second of two field theory
techniques for performing an RG\cite{dg}.  Within this procedure, one
needs only to keep two terms in the RG, the value of $\sigma_0^2$ and
the value of $S$, as these are the only two relevant terms.  It is possible,
of course, that higher loop corrections will make other operators relevant,
and destabilize the fixed point; this is something that cannot be analyzed
within this approach.  However, this approach will yield results that
agree with the Wilson-Fisher technique so long as the fixed point is
stable.  Let me emphasize this point: if no other operators become relevant,
then this technique agrees with the Wilson-Fisher procedure to
all orders.   If other operators become relevant, then the procedure does
not work and the fixed point has more than one unstable direction.  
What we show in this section is that if no other operators become
relevant and the fixed point remains stable then the exponents are unchanged
to all orders.  What was shown in the previous section is that there is
a stable fixed point to lowest order for $D-d<2$, with 
exponents as given above.
It is reasonable to assume that the fixed point remains stable to all
orders, at least for some range of $\epsilon$, so that the exponents remain
unchanged to all orders at this fixed point.

In the case of the one loop calculation above, it was found that
the vertex renormalization for $\lambda$, the vertex 
renormalization for $\sigma^2$,
and the renormalization of $\omega^2$ all had the same coefficient.  This will
remain true to all orders.  This is due to a Ward identity, explained at the
end of this section.  
Therefore, the general RG equation to all orders will take the form
\be
(1-\frac{\delta\Lambda}{\Lambda}F_1(S,D,d)))
\sigma^2(x)=\int d^{(D-d)}\omega \,\langle x,t=0|
\Bigl(-(1+\frac{\delta\Lambda}{\Lambda}F_2(S,D,d))
\partial_{x}^2+
(1+\frac{\delta\Lambda}{\Lambda}F_1(S,D,d))
(\omega^2
+\lambda(x))\Bigr)^{-1})|x,t=0\rangle
\ee
where $F_1,F_2$ are some generic functions of disorder strength and dimensionality.
By the same procedure of rescaling, and writing $\sigma^2=\sigma_0^2+\delta\sigma^2$,
as before, we can obtain
\be
\frac{d{\rm ln}\sigma_0^2}{d{\rm ln}\Lambda}=D-2-(F_1(S,D,d)-F_2(S,D,d))(1-\frac{D-d}{2})
\ee
\be
\frac{1}{2}\frac{d{\rm ln}S}{d{\rm ln}\Lambda}=D-2-(F_1(S,D,d)-F_2(S,D,d))
(1-\frac{D-d}{2})
-d/2
\ee
Then, although the fixed point of the RG flow will be at a different value of
$S$ than was obtained to lowest order, the exponents will be unchanged
from their lowest order values.

Let us consider why this Ward identity holds.  First, consider
the equivalence between
the renormalization of $\sigma^2$ on the left-hand side of the self-consistency
equation and the renormalization of $\lambda$ on the right-hand side.  It
is apparent that both renormalizations are vertex renormalizations and are
the same by definition.  

The more interesting aspect of the Ward identity
is the equivalence between the renormalization of $\sigma^2$ and the 
renormalization of the $\omega^2$ term.  
This, however, is again almost true by definition.  The term $\omega^2$ is
constant in space, and a function only of frequency.  The Green's function
for given frequency $\omega$
is constant under a shift $\omega^2\rightarrow\omega^2+\Delta$ and
$\lambda\rightarrow\lambda-\Delta$, for some constant $\Delta$.  
For this property to hold under renormalization, we need the desired Ward
identity.  Alternatively we may say that for any given frequency $\omega$, the
term $\omega^2$ plays exactly the same role in the Green's function as a
constant term in $\lambda$ would, and so the renormalizations must be equal.

For a more diagrammatic derivation of the identity, proceed as follows.
The coefficient of the $\omega^2$ term can be obtained
by differentiating the inverse of the average Green's function with respect to
$\omega^2$ at $\omega=0$.  In any given diagram for the average Green's
function, one can differentiate the inverse of
any one of the propagators in the diagram
with respect to $\omega^2$.  This gives one a sum over different places to
insert the derivative.  However, each place one inserts gives a result
for the diagram which is equivalent to placing a scattering vertex for 
$\lambda$ at that point, and thus the total renormalization of the $\omega^2$
vertex is the same as the renormalization of the $\lambda$ vertex.  For 
example,
in figure 2 for the one-loop self-energy, the renormalization of the $\omega^2$
term can be obtained by differentiating the inverse of the propagator in
the loop with respect to $\omega^2$.  However, this yields exactly the diagram
in figure 3.

The existence of this Ward identity does not require the formulation of the
RG used in this section.  A similar identity would exist within a 
Wilson-Fisher renormalization group.  This would then connect the 
renormalization of the $\omega^2$ term to that of the $\sigma^2$ and $\lambda$
terms at lowest order as found in the last section.  
At higher orders, other terms would also be connected.
For example, the existence of the symmetry would also require
that a term such as $A(x)\omega^2$, that is, a spatially fluctuating
$\omega^2$ term, would have a renormalization connected to that of
a term $A(x)\lambda$, and a term $A(x)\sigma^2$.  This last term
would imply a spatially varying magnitude of fluctuations in the disorder 
$\delta\sigma^2$, which would mean a non-Gaussian distribution of the
disorder.  From the Ward
identity, we would still have a connection between the renormalization of
the $\omega^2$, $\sigma^2$ and $\lambda$ terms, which would be enough to
obtain the desired results for critical exponents, to all orders.

The only thing the Ward identity does not guarantee is the stability and
existence of the fixed point.  It only guarantees the exponents if the
fixed point exists and is stable.  As $D-d\rightarrow 2$, the fixed
point runs to stronger and stronger disorder.
This will be discussed more in the conclusion.
\section{A Heuristic Treatment of the RG}
We present a simple alternative treatment of the problem which supports the
results of the RG and provides a physical motivation for it.
First, let us present an alternative version of the
Harris criterion.  For the pure system, at criticality, the self-consistency
equation is given by
\be
\label{csc}
\sigma^2=\int^{k^2+\omega^2<\Lambda^2} d^dk \,d^{(D-d)}\omega \,\frac{1}{k^2+\omega^2}
\ee
where we have now inserted a cutoff for high energy states of the field
$\phi$.
We find by doing the integral in equation (\ref{csc}) 
that $\sigma^2 \propto \Lambda^{(D-2)}$.
However, if we consider a disordered system, we find that, after
removing states with $k^2+\omega^2>\Lambda^2$, we are considering a case
in which we have smeared out space over a length of order $1/\Lambda$, which
corresponds to a $d$-dimensional volume $V$ of order $1/\Lambda^d$.
The average value of the fluctuation in $\sigma^2$ over 
this volume is of order
$V^{-1/2}$, or $\Lambda^{d/2}$.  When the Harris criterion indicates that
disorder is relevant, we find that $\Lambda^{d/2}>\Lambda^{(D-2)}$ for
small $\Lambda$.  This is to be interpreted as saying that there are not
enough low energy states in the pure system to produce the required $\sigma^2$.

However, we expect that disorder will increase the low energy density of states,
and we heuristically modify the Green's function to
$\frac{1}{k^a+\omega^2}$, with $a>2$, and we change the cutoff to
$k^a+\omega^2<\Lambda^2$ so that we have
\be
\label{dsc}
\sigma^2=\int^{k^a+\omega^2<\Lambda^2} d^dk \, d^{(D-d)}\omega \, \frac{1}{k^a+\omega^2}
\ee
In this case, we find by doing the integrals that
$\sigma^2\propto \Lambda^{2d/a-2-d+D}$.  However, we are now smearing the
system out over a length of order $1/\Lambda^{(2/a)}$.  For small $\Lambda$,
this is less the length $1/\Lambda$ which we had in the pure system.  
This is to be expected, indicating the motion
has become subdiffusive.  This leads to an average of $\sigma^2$ over
the $d$-dimensional volume of order $\Lambda^{(d/a)}$.  Equating these
two results for $\sigma^2$, one given by the integral in equation
(\ref{dsc}) and the other given by the volume average of $\sigma^2$, we find that
\be
2d/a-2-d+D=d/a
\ee
or $a=d/(2+d-D)$.  This agrees with the value of $a$ obtained by the
RG treatment of the previous section.  Further, if we imagine slightly
leaving the critical point, we can imagine that the self-consistency equation
gets replaced by 
$\sigma^2=\frac{1}{k^a+\omega^2+\lambda}$, 
where $\lambda$ and $\sigma^2$ are now taken spatially constant.
If we slightly adjust $\sigma^2$ away from the critical value, we can calculate
the change in $\lambda$, and thus the correlation length of the system,
and we find that we obtain $\nu d=2$. 
\section{The Zero Energy Wave Function}
As we increase $\sigma^2_0$, with fixed randomness $\delta\sigma^2$, we find
a sequence of different phases.  Assume we are given $\lambda(x)$ as
a function of $\sigma^2_0+\delta\sigma^2$.  
For small $\sigma^2_0$, the operator $(-\partial_{\mu}^2+\lambda(x))$ has a
gap, and there are no eigenstates of this operator with eigenvalues
below the gap. 
For the case of the zero temperature quantum phase transition, where $D=d+1$,
this gap corresponds to an energy gap for the system.
This is the Mott insulator phase discussed in the introduction. 
As $\sigma^2_0$ is increased the gap decreases and at a certain point
the system enters
a Griffiths phase.  In this case, there are eigenstates at 
arbitrarily small energy,
all of which are localized, and the average correlation function decays 
exponentially.
At the critical $\sigma^2_0$, the average correlation function 
acquires power law
behavior, and there appears a state at zero energy which is delocalized.
Above the critical $\sigma^2_0$, particles begin to condense into this
state.  We would like to examine the wave function $\alpha(x)$ of this state.

We will first give a physical argument for the form of the wave function.
Then, we will derive this to first order in $\epsilon$ through the RG.  
Unfortunately, this result cannot be derived to all orders in $\epsilon$.

Arguing physically, the following should be a good description
of this wave function.  Consider a problem at short distances.
From the RG, we know that, on a short scale, one can obtain $\lambda$
directly from $\sigma^2$.  So, the wave equation,
$(-\partial_x^2+\lambda)\alpha(x)=0$, can be solved approximately
on a short scale knowing only the local fluctuations in $\sigma^2$.
However, the wave equation is homogeneous, and defines
$\alpha(x)$ only up to a multiplicative constant.  So, the zero energy
wavefunction in some region in space is defined by the local disorder,
up to a multiplicative constant.  This constant will be
set by the behavior of the disorder on larger length scales.  So if we want
to find the value of $\alpha(x)$ at some point, we proceed as follows:
consider short distance fluctuations to obtain $\alpha(x)$ as a solution
of a wave equation on some short scale.  The overall multiplicative constant
on this wave function is not known.  So, $\alpha(x)$ at a given point
is known up to a multiplicative constant set by longer scales.  Solving
at a slightly longer length scale, we can find $\alpha(x)$ up to a constant
set at even longer length scales.  

So, we would find that $\alpha(x)$ is given by a product of
these multiplicative constants on larger and larger length scales, where
these constants will be drawn from some random distribution.
This is simply a statement that the zero energy wavefunction, or the
value of the condensate, is a multiplicatively renormalizable operator, which
gets a random multiplicative renormalization at each step.
Thus, $\alpha(x)$ would have a log-normal distribution.
Assuming scale invariance we would find that 
\be
\alpha(x)=e^{\beta(x)}
\ee
where, in order to produce the same fluctuations on all length scales,
 $\beta$ has a Gaussian distribution with
\be
\label{bpd}
\langle \beta(p)\beta(q)\rangle=\frac{g}{p^d}\delta(p+q)
\ee
where $g$ is some constant measuring the strength of the disorder.
In two space dimensions, we can give $\beta$ the probability distribution
\be
e^{-\frac{1}{g}\int d^2x (\partial_x \beta(x))^2}
\ee

Let us now derive this result from the RG.  The zero energy wavefunction
$\alpha(x)$ can be obtained by considering the correlation function
of $\phi(x,t)$ and $\overline\phi(y,t')$ where $y$ is some point very far
from $x$.  So, we must look at how the correlation function 
$G(x,t;y,t')$ renormalizes
under the RG, beyond the simple calculation of the average correlation
function in the previous sections.  The interesting part is the
renormalization of $\phi(x)$.  Let us look at this problem using
moments.  That is, we will look at the average over disorder of
the $n$-th moment of $G(x,t;y,t')$.  Let us start with the second
moment for simplicity.  This is an average of a product of two Green's 
functions.

Consider what we may call a renormalization of a vertex.   
We must have both Green's functions starting at the same point, $x,t$.
One may connect the two lines with a single scattering off of $\lambda$,
with momentum of order $\Lambda$ running around the loop, with low
momentum leaving the diagram.  See figure 4 for the appropriate diagram.
The result of this is that the vertex is renormalized under RG flow as
\be
\frac{d{\rm ln}V_2}{d{\rm ln}\Lambda}=c_3L=c_3c_1^2\Lambda^{8-2D}S
\ee
where $V_2$ is the vertex.

In general, for the $n$-th moment, there are $\frac{n(n-1)}{2}$
ways to connect the lines at the vertex, and so we find
\be
\label{mrn}
\frac{d{\rm ln}V_n}{d{\rm ln}\Lambda}=\frac{n(n-1)}{2}c_3L=
\frac{n(n-1)}{2}c_3c_1^2\Lambda^{8-2D}S
\ee
Interpreting this result, we find that the $n$-th moment of $\alpha(x)$
at some point can be expressed, up to the multiplicative renormalization
calculated here, as the $n$-th moment of $\alpha(x)$ at that point with
a smoothed disorder potential.  Looking at the renormalization, though,
we see that it is exactly the form expected for a log-normal distribution
of $\alpha(x)$.  That is, if we took $\alpha(x)$ at some point to be expressed
as a random multiplicative renormalization of the value of $\alpha(x)$ at 
that point calculated from a smoothed disorder potential, we would obtain
a result of the form of equation (\ref{mrn}).

Let us show that the moments are those expected from a log-normal
distribution, as well as obtaining the value of $g$, measuring fluctuations
in $\beta$, as defined above.  Using $\alpha(x)=e^{\beta(x)}$,
computing the average of $\alpha^n(x)$, over fluctuations in $\beta$
down to some scale $\Lambda$, we find
\be
\langle \alpha^n(x) \rangle=\int [d\beta(p)]e^{\int_{\Lambda}
d^dp \, (n\beta(p)-\frac{p^d}{2g}
\beta(p)^2)}
\ee
This is equal to
\be
e^{\int_{\Lambda} d^dp \frac{gn^2}{2p^d}}
=
e^{2 \frac{\pi^{d/2}}{\Gamma(d/2)} \int_{\Lambda} d{\rm ln}{p} \frac{gn^2}{2}}
\ee
Finally, we see that 
\be
\label{mffl}
\frac{d{\rm ln}(\langle \alpha^n\rangle)}{d{\rm ln} \Lambda}=
\frac{\pi^{d/2}}{\Gamma(d/2)} gn^2
\ee
So we find, comparing equations (\ref{mrn}) and (\ref{mffl}), that
\be
g=\frac{c_3 L \Gamma(d/2)}{2 \pi^{d/2}}
\ee
where we need to choose $g$ to make the coefficients of the
$n^2$ term in equations (\ref{mrn}) and (\ref{mffl}) the same.
The term in $n$ can be different, as this simply represent an overall
scale for the wavefunction.

Given this form of the wave function, with log-normal fluctuations, 
the change in the low energy
density of states should not be a surprise.  Originally, the field
$\phi$ had the action
\be
\int d^dx \,d^{(D-d)}t\,\Bigl(
-\partial_{\mu} \overline \phi(x,t) \partial^{\mu} \phi(x,t) 
+\lambda(x) \overline \phi \phi \Bigr)
\ee
Given the zero energy wave function, we can write this as
\be
-\int d^dx \, d^{(D-d)}t \, \alpha(x)^2 \partial_{\mu} 
(\alpha^{-1}(x)\overline \phi(x,t)) \partial^{\mu}(\alpha^{-1}(x) \phi(x,t))
\ee
This leads to a random stiffness problem, which is related to the
problem of disordered SUSY quantum mechanics\cite{SUSY} and the problem
of Dirac fermions in a random vector potential\cite{Mudry}.  Both of these
problems are known to have an increase in the low energy density of
states.  In particular, for the Dirac problem in which there exists
a dimensionless measure of disorder, the low energy density of states is
a power law with continuously variably exponent.
\section{Conclusion}
In conclusion, we have presented a renormalization group treatment of
the disordered large $N$ sigma model.  We have calculated the exponents
describing the average correlation functions.
However, there is still
much that we would like to know.

First, we have not considered the behavior of averages of higher moments
of the correlation functions, or typical behavior of the correlation
functions.  These will be discussed in another publication\cite{me2}.
The calculation of these higher moments is closely related to the
calculation of the zero energy wave function.  It is shown\cite{me2} that
\be
\langle G^n(p,\omega) \rangle \propto \langle G(p,\omega) \rangle^n
p^{-n(n-1)c_3L}
\ee
similar to the result for the wave function derived above.  Here, $c_3$
and $L$ are the constants defined in section II, while the expectation
value is an average of the $n$-th moment of the correlation function.

Also, there exists another RG treatment of the same system\cite{xiaogang},
based on a Callan-Symanzik type RG for the large $N$ system.
In that work, no perturbatively accessible fixed point was found, and I
do not fully understand why their technique fails.  Of course, this
work\cite{xiaogang} is not necessarily in contradiction with the results
obtained here.  Since no fixed point was found with their technique, this
may simply be failure of technique, rather than indicating different
results for the same system.

Let us consider the differences in techniques used in more detail.
It is conceivable that the lack of a perturbatively accessible fixed 
point in that work\cite{xiaogang} is due to their use of a Callan-Symanzik 
technique, which is equivalent to the Wilson-Fisher RG only for a field 
theory which possesses a renormalizable continuum description.  However, 
the $\delta$-function interaction is non-renormalizable as a continuum 
theory (in that work a quartic interaction, which is renormalizable, was 
used, but the technique of summing polarization bubbles used there made the 
results equivalent to the $\delta$-function theory) and the inverse 
polarization bubble used to calculate fluctuations in $\lambda$ from 
fluctuations in $\sigma^2$ grows rapidly in the ultraviolet which can lead to 
poor behavior of the diagrammatic expansion of the continuum theory.  
A perturbative RG implies an ordering by momenta, in which high momentum
processes are calculated before low momentum processes.  For the expansion
to be valid, this ordering by momenta must be correct, in that corrections
to high momentum processes due to low momentum effects must be small.  I
have checked this for my procedure; I am not aware if it is true for other
techniques on the same problem.

Although results have been obtained for average quantities to all orders,
one still may inquire about the radius of convergence of the
expansion.  It is apparent that if $(D-d)\geq 2$ then 
the dynamic critical exponent $z$ resulting from equation (\ref{dce}) becomes
infinite.  When $(D-d)>2$, however, it is
possible for the system to do a phase transition in a finite region in
the $d$ dimensional space, since there is enough volume in the remaining
$D-d$ dimensions.  Thus, certainly when $D-d=2$, and possibly before, the
perturbation theory must break down.  Further, results for the
higher moments of correlation functions may have interesting behavior
to higher order.

Also, we would like to understand $1/N$ corrections to this problem.
To lowest order in $1/N$ and $\epsilon$
such corrections only modify the
RG equations by changing the scaling dimension of $\sigma^2$\cite{1N} in the
pure system, and thus
changing the equation to
\be
\frac{d{\rm ln}\sigma_0^2}{d{\rm ln}\Lambda}=
D-2+\eta
-(c_3-c_2)c_1^2\Lambda^{8-2D}S(1-\frac{D-d}{2})
\ee
\be
\frac{1}{2}\frac{d{\rm ln}S}{d{\rm ln}\Lambda}=D-2+\eta
-(c_3-c_2)c_1^2\Lambda^{8-2D}S
(1-\frac{D-d}{2})
-d/2
\ee
where in the physically interesting case of $d=2, D=3$ we find
$\eta=\frac{32}{3\pi^2}\frac{1}{2N}$.
Then, we would expect to still find $\nu d =2$ for the disordered 
critical point.
\section{Acknowledgements}
I would like to thank David Huse for many useful discussions on disordered
quantum critical points.  I would also like to thank Shivaji Sondhi and 
Andy Green.

\newpage
\begin{figure}[!t]
\begin{center}
\leavevmode
\epsfig{figure=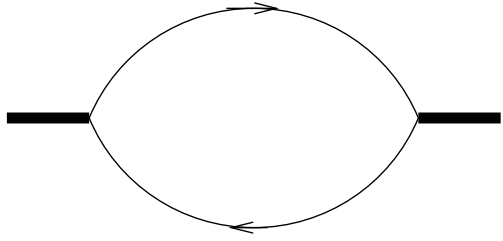,height=3cm,angle=0}
\end{center}
\caption{Polarization bubble.  Thick lines represent either scattering vertex
off $\lambda$ or scattering vertex used to define $\sigma^2$ in
self-consistency equation.}
\label{fig1}
\end{figure}
\begin{figure}[!t]
\begin{center}
\leavevmode
\epsfig{figure=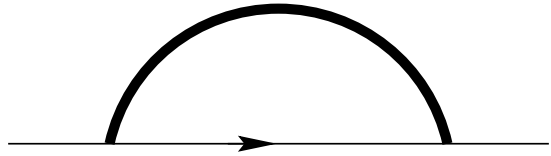,height=2cm,angle=0}
\end{center}
\caption{Self-energy correction due to fluctuations in $\lambda$.  Joining the
thick lines in a loop denotes averaging $\lambda$ over disorder in
$\sigma^2$. Momentum of order $\Lambda$ flows around loop.}
\label{fig2}
\end{figure}
\begin{figure}[!t]
\begin{center}
\leavevmode
\epsfig{figure=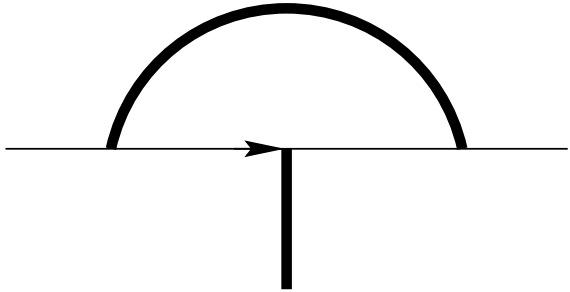,height=4cm,angle=0}
\end{center}
\caption{Vertex correction due to fluctuations in $\lambda$.  This represents
both renormalization of vertex defining scattering off of $\lambda$ and
renormalization of vertex defining $\sigma^2$ in self-consistency equation.}
\label{fig3}
\end{figure}
\newpage
\begin{figure}[!t]
\begin{center}
\leavevmode
\epsfig{figure=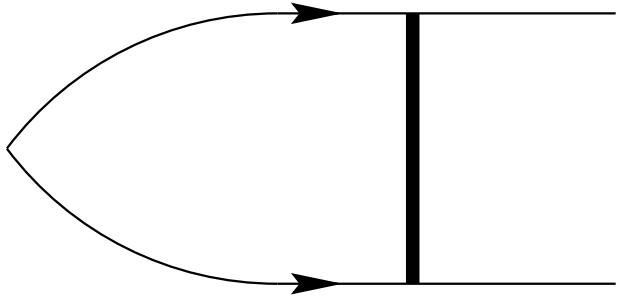,height=4cm,angle=0}
\end{center}
\caption{Renormalization of vertex in computing higher moments of
Green's function.  Two Green's functions start at the same point.  After
Fourier transforming, this implies that they start with given
total momentum.  By including fluctuations in $\lambda$, with momentum
of order $\Lambda$ running around the loop, one can define a renormalized
vertex.}
\label{fig4}
\end{figure}
\end{document}